\documentstyle[12pt]{article}

\renewcommand{\arraystretch}{1.5}

     \def\BE{\begin{equation}}
     \def\EE{\end{equation}}
     \def\BEA{\begin{eqnarray}}
     \def\EEA{\end{eqnarray}}
     \def\BA{\begin{array}}
     \def\EA{\end{array}}

\newenvironment{myquote}{\begin{quote}\small\it\baselineskip2pt
  }{\end{quote}}

     \def\BQ{\begin{myquote}}
     \def\EQ{\end{myquote}}

\def\ds{\partial\!\!\! /}

\def\ol#1{\overline{#1}}
\def\wt{\widetilde}

\def\1{\hbox{\bf 1}}

\def\de{\delta}

\def\N{{{\cal N}}}

\def\1{\hbox{\bf 1}}

\def\ol#1{\overline{#1}}
\def\pr{^\prime}

\def\tt{\otimes}

\def\R{\widehat R}
\def\1{\hbox{\bf 1}}

\def\al{\alpha}
\def\be{\beta}

\def\xxpp{(x\!\cdot\! x)(\pi\!\cdot\! \pi)}
\def\xx{\:(x\!\cdot\! x)\:}
\def\dd{\:(\partial \!\cdot\! \partial )\:}
\def\XX{\:(\xi\!\cdot\! \xi)\:}
\def\pp{\:(\pi\!\cdot\! \pi)\:}
\def\xp{\:(x\!\cdot\! \pi)\:}
\def\xpn(#1){\:(x\!\cdot\! \pi)^{#1}\:}
\def\ad{\:(\xi\!\cdot\! \partial )\:}
\def\ap{\:(\xi\!\cdot\! \pi )\:}
\def\ax{\:(\xi\!\cdot\! x )\:}

\def\xd{\:(x\!\cdot\! \partial )\:}
\def\i{\hbox{i\,}}
\def\ds{\displaystyle}
\def\exp{\hbox{exp}}

\begin{document}


\renewcommand{\thefootnote}{\fnsymbol{footnote}}
\renewcommand{\footnoterule}{\rule{5cm}{0.5pt}}


\hfill LBL--36151\par
\hfill UCB-PTH--94/24\par
\hfill July 1994\par
\vspace*{40pt}
\begin{center}
{\LARGE
Generalized {\sl q}-exponentials related to orthogonal\\
quantum groups and Fourier transformations\bigskip\\
of noncommutative spaces\footnote{Work supported by the Fritz Thyssen
Foundation.}}\footnote{This work was supported by part by the
Director, Office of Energy Research, Office of High Energy and
Nuclear Physics, Division of High Energy Physics of the U.\ S.\
Department of Energy under Contract DE-AC03-76SF00098 and
by part by the National Science Foundation under grant
PHY90-21139.}\\
\vspace{30pt}
{\bf Arne Schirrmacher \footnote{ e-mail:
\sf schirrmacher@theorm.lbl.gov.}} \\
\vspace{20pt}
{\it Department of Physics\\
and\\
Theoretical Physics Group\\
Lawrence Berkeley Laboratory\\
University of California\\
Berkeley California 94720}\\
\end{center}
\par
\parindent0pt
\vspace{100pt}

\begin{quote}
{\bf Abstract}\\
An essential prerequisite for the study
of $q$-deformed physics are particle
states in position and momentum representation.
In order to relate $x$- and
$p$-space by Fourier transformations the appropriate
$q$-exponential series
related to orthogonal quantum symmetries is constructed.
It turns out to be a
new $q$-special function giving rise to $q$-plane wave
solutions that transform
with a noncommutative phase under translations.
\end{quote}
\thispagestyle{empty}
\eject
\setcounter{page}{1}

\renewcommand{\thefootnote}{\arabic{footnote}}
\setcounter{footnote}{0}


\thispagestyle{empty}
\mbox{ }

\vskip 1in

\begin{center}
{\bf Disclaimer}
\end{center}
\vskip .2in
\begin{scriptsize}
\begin{quotation}
This document was prepared as an account of work sponsored by the
United States
Government. While this document is believed to contain correct
information, neither
the United States Government nor any agency thereof, nor The
Regents of the University
of California, nor any of their employees, makes any warranty,
express or implied,
or assumes any legal liability or
responsibility for the accuracy, completeness, or usefulness
of any information, apparatus, product, or process disclosed,
or represents that its use
would not infringe privately owned rights.  Reference herein
to any specific commercial
products process, or service by its trade name, trademark,
manufacturer, or otherwise,
does not necessarily constitute or imply its endorsement,
recommendation, or favoring by
the United States Government or any agency thereof, or The
Regents of the University of California.
The views and opinions of authors expressed herein do not
necessarily state or reflect those of the United States Government or
any agency thereof of
The Regents of the University of California and shall not be used for
advertising or product endorsement
purposes.
\end{quotation}
\end{scriptsize}

\vskip 2in

\begin{center}
\begin{small}
{\it Lawrence Berkeley Laboratory is an equal opportunity employer.}
\end{small}
\end{center}
\newpage
\renewcommand{\thepage}{\arabic{page}}
\setcounter{page}{1}


\section{Introduction}

The theory of quantum groups is closely related to the structure
of quantum mechanics as
commutative coordinate algebras are replaced by noncommutative
operator algebras, that are
in a strict mathematical sense deformations of a commutative
structure.
The developments in noncommutative geometry and quantum groups
provide new motivation to
address the question of quantized spacetime, that have been
discussed almost as long as
quantum mechanics itself \cite{Heis}. When the angular momentum
algebra becomes a noncommutative
operator algebra in quantum mechanics yielding the discretization of
energy levels, is it then not
also conceivable that noncommutative position and momentum algebras
give rise to a quantization or
discretization of the spacetime continuum itself?
This idea was discussed some decades ago when divergence problems in
quantum field theory were
traced back to the probably unphysical use of the continuum \cite{Sny}.

{}From a current point of view it is commonly accepted that
at some small scale the smooth
manifold picture of spacetime breaks down. Pictures of a
foam-like structures \cite{Wheeler}
or of a thermodynamic nature \cite{Zimmermann} of spacetime
were evoked.
Clearly, the questions addressed here are outside experimental
accessibility as measurability
becomes impossible long before a quantum nature of spacetime
 shows up \cite{WS}. For a fundamental
understanding of forces and matter, however, a consistent
mathematical description cannot be
abandoned and the obvious unsolved problems of quantum physics
with gravity and the submicroscopic
structure of spacetime have to be approached.

We foster the idea that extending the basic quantum mechanical
paradigm well-known for the quantization
of angular momentum to spacetime observables themselves should be
considered a natural
approach to the microstructure of spacetime.
Current research on quantum groups as the mathematical
structure for such a
construction focuses on analysing $q$-deformed structures
of quantum theory.
In this article we
concentrate on the important feature of Fourier transformations
that relate noncommutative position and momentum spaces. The basic
ingredient here are
appropriate generalized
$q$-exponential series that are related to orthogonal quantum
groups with $\R$ operators satisfying the Birman-Wenzel-Murakami
algebra.
This will
be the central issue of this article. From a purely
mathematical point of
view the results may
also be of interest as they provide a new class of
$q$-special functions.

For the so-called quantum plane (with $\R$ of
Hecke type) the appropriate exponential
has been given at the beginning of the
century by F. H. Jackson (without any relation to
quantum groups of course) and noncommutative eigenvalues and
Fourier transformations were
introduced recently \cite{CZ}, which is a major starting point
for this work.
In this approach we study first the quantum symmetry structure on
the noncommutative
algebraic level as far as possible and discuss eigenvalue problems
in the noncommutative setting.
In a second step the usual quantum mechanical construction of Hilbert
space representations
of (essentially) self-adjoint operators should be pursued.

 It should be mentioned that a sort of generalized $q$-exponentials
have already been considered
in \cite{Maj} for general $\R$-matrices. The results, however, exclude
$q$-deformed Euclidean
and Minkowskian spaces and are only applicable for freely generated
algebras, that have so far no
application in physical theory.
\medskip\par

The paper is organized as follows: In section 2 the basic
$\R$-matrix properties for orthogonal
quantum groups and their differential calculus are given. This
algebraic structure is
augmented by quantum spaces of finite displacement in position
and momentum. The generators
of translations are identified and their commutation relations
with the quantum spaces calculated.
In section 3 the generators of translation are integrated
up to finite translation operators,
this involves new generalized $q$-exponentials.
For this construction new $q$-combinatorial
coefficients have to be introduced also.
Then the actions of the translation operators in
position and momentum space on functions
of conjugate variables is analysed. Some technical
material is collected in the appendix.
In section 4 momentum eigenstates are introduced and some of their
properties discussed.
Employing the idea of an average or integral of
noncommutative functions from \cite{WZ}
Fourier transformations can be
defined in the same manner as for the quantum planes \cite{CZ}.
The article concludes with a discussion of open
problems and some comments on related work.

\bigskip

\section{Euclidean and Minkowskian quantum
vectors with finite translations}

Let us start with reviewing some basic properties of orthogonal
quantum groups in general and
introducing additional quantum spaces of finite displacements for
this case following \cite{CZ}.
We adopt the following
{\it conventions}: The normalization of $\R$-matrices has been chosen
differently by different
authors and there is also frequent usage of $q^{-1}$ instead of $q$.
In this article we use $q$
as in \cite{FRT} but normalize the $\R$-matrix such that $-1$ is the
eigenvalue related to the
antisymmetric part instead of $-q^{-1}$ in \cite{FRT}, consequently
all commutation relations
given in that paper apply unaltered to our case. The normalization
we adopt agrees with the
treatment of the Lorentz group in \cite{OSWZ} but one has to
replace there $q$ with $q^{-1}$ for comparison.
In writing $\R$-matrix relations without indices it is assumed
that all quantum vectors carry upper indices
and $\R$ and other matrices are of type $M^{ij}_{\ \ kl}$.
\medskip\par

The $\R$-matrices of orthogonal quantum groups in $N$ dimensions
satisfy in addition to the braid
group relation (Yang-Baxter equation) the characteristic equation
\BE
(\R-q^2)(\R-q^{-(N-2)})(\R +1) = 0
\EE
making it a Birman-Wenzel-Murakami algebra \cite{BWM}.
This corresponds to a projector decomposition into a
traceless symmetric, an antisymmetric, and a
trace projection:
\BE
\R=q^2P_S+q^{-(N-2)}P_T-P_A
\EE

The trace projection can be written in terms of the metric
\BE
P_T=\frac{1}{\N}\ g\tt g
\EE
with
\BE\BA{rcll}
g_{ij}=\de_{i,N+1-j}\ \rho_j\ ,&\quad&
\rho=(q^{\frac{1-N}{2}},q^{\frac{3-N}{2}},...,
q^{-\frac{1}{2}},1,q^{\frac{1}{2}},...,
q^{\frac{N-3}{2}},q^{\frac{N-1}{2}})&\hbox{for N odd}\\
&&
\rho=(q^{1-N},q^{2-N},...,q^{-1},1,1,q,...,
q^{N-2},q^{N-1})&\hbox{for N even\ .}
\EA
\EE
The normalization factor $\N$ has to be chosen such that $P_T^2=P_T$.
Hence
\BE
\N=g^{ij}g_{ij}=1+\sum_{k=1}^{N-1}q^{2(k-\frac{N}{2})} =
1+\frac{q^{N-1}-q^{1-N}}{q-q^{-1}}\ .
\EE
The metric satisfies some very useful relations with the $\R$-matrix
\cite{FRT}, that allow
us to `move' the braid group operator from one pair of variables to
another, which amount to
a rotation of the $\R$-matrix in the pictorial representation of the
Birman-Wenzel-Murakami algebra.
\BE
{\R^{-1}}^{ij}_{\ \ kl}= q^{-2}g^{ia}\R^{jb}_{\ \ ak}g_{bl}
=q^{-2}g_{ka}\R^{ai}_{\ \ lb}g^{bj}
\EE
$\R$ can be replaced with its inverse using the metric:
\BE
\R=q^2 \R^{-1}+(q^2-1)\1 -(q^2-1)\  g\tt g
\EE
Due to the fact that in the $\R$-matrix the eigenvalues of the two
symmetric projections are different,
it is convenient to introduce a matrix $\R\pr$ that in general, however,
is no representation of the
braid group
\BE
\R\pr=
P_S+P_T-\al(q) P_A
\EE
with $\lim_{q\to1}\al(q)=1$, that can be expressed as a
polynomial in
$\R$, which satisfies a simple
relation with $\R$ \cite{FRT} needed in the following
\BE
(\R+1)(\R\pr-1)=0\ .
\EE
It is this splitting in the different eigenvalues of the
traceless symmetric and the trace
projection in particular that accounts for the major
difference to the case of the quantum
plane where the well-known Jackson exponential can
be used.\medskip\par

In the special case of four dimensions there exists a
further decomposition of the antisymmetric
projection into selfdual and anti-selfdual
parts $P_A=P_++P_-$ due to the possibility to introduce
spinors. Then one finds two solutions of the Yang-Baxter equation:
\BE
\BA{rclrcl}
\R_I   &=& P_S+P_T-q^{-2}P_+-q^2P_-}\cr
\R_{II} &=& q^2P_S+q^{-2}P_T-P_+-P_-
\EA
\EE
These two $\R$-matrices were found for the Lorentz group in
\cite{OSWZ} first.
Since its projectors are related to those of the known
deformations of the $SO(4)$
by a similarity transformation (Drinfeld twist) this
property applies to all known
orthogonal four-dimensional quantum groups. For the
mentioned deformation of the
 Lorentz group the metric reads
\BE
g=\pmatrix{
0&q^{-2}&0 &0 \cr
1 &0&0&0\cr
0&0&0&-1\cr
0&0&-1&q^{-2}-1
}
\EE
giving the same $\N$ as equation (5). As a consequence the quantum
Lorentz groups
are included in the general treatment of orthogonal quantum groups
in $N$ dimension.\medskip\par

The commutation relations for quantum vectors $z$ which coordinates are
symmetric can be given as
\BE
zz=(P_S+P_T)zz=\R\pr zz
\EE
or
\BE\BA{rclrcl}
zz &=&\left(q^{-2}\R+(1-q^{-N})P_T\right)\, zz
&=& \left(q^{-2}\R+\wt\mu\ g\tt g\right)\, zz  \\
   &=&\left(q^{2}\R^{-1}+(1-q^{N})P_T\right)\,
zz &=&\left(q^{2}\R^{-1}+\mu\ g\tt g\right)\, zz
\EA
\EE
with
\BE
\mu=(1-q^N)/\N=\frac{1-q^2}{1+q^{-(N-2)}}\ ,\qquad
\wt\mu=(1-q^{-N})/\N=\frac{1-q^{-2}}{1+q^{(N-2)}}\ .
\EE
The commutation relations for a quantum vector transformed by
a quantum  orthogonal transformation $z\to Tz$
with $\R\,TT=TT\R$ and $g\,TT=g$ remain
the same due to (2) and (8) and the length $g\,zz$ is invariant.
\medskip\par

In the following we use four such quantum vectors: the position
operator
$x$ and a finite displacement of position $\xi$
and the momentum operator $p=\i \partial $ and a finite displacement
of momentum $\pi$. We have to specify a set of consistent
commutation relations among these four quantum planes.
The relation between coordinates and derivatives is
\BE
\partial x= g + q^2\R^{-1}\ x\partial
\EE

 in order to have linear relations compatible with the $xx$-relations
\cite{WZ}. It is also possible to use $q^{-2}\R$ replacing
$q^2\R^{-1}$ in (15) as there are always two
choices to define the differential calculus
on quantum spaces.\footnote{As discussed below,
both relations are realized
if one wants to introduce a
$*$-structure: e.g.\ taking $\partial $ antihermitean we get
$\partial \ol{x}= g + q^{-2}\R\ \ol{x}\partial $.}
The requirement that translations in position and
momentum space should yield new quantum
planes $x\pr=x+\xi$ and $p\pr=p+\pi$ gives
\BE\BA{rcl}
\partial \pi &=&\R^{-1}\pi\partial \\
\xi x &=&\R^{-1}x \xi
\EA\EE
 since using (11)
\BE\BA{rcl}
x\pr x\pr &=& (x+\xi)(x+\xi)  \\
 &=& xx+\xi\xi+x\xi+\xi x  \\
  &=& \R\pr (x+\xi)(x+\xi) -(\R\pr-1)(x\xi+\xi x)
\EA
\EE
where the last term vanishes now due to (16) and (8). Here
again $\R$ could replace
$\R^{-1}$ in (16) (independently from the above choice) but
using (15) it turns out
that only for the choice (16) we will find an appropriate
generator of translations
in the algebra.

As a consequence of (15) and (16) we get
\BE
\partial \xi=q^{-2}\R \xi\partial \ .
\EE
And finally the translated quantum vectors should have
the same relations with the other
finite displacement parameters:
\BE
\BA{rclrcl}
\pi x &=& q^{-2}\R x\pi  \\
\pi  \xi &=& q^{-2}\R \xi \pi
\EA
\EE
This fixes all commutation relations. As mentioned in \cite{CZ}
this structure can be interpreted as a braided quantum
space with differential structure in the sense
of \cite{braid}. We do not adopt this more restrictive
framework here and prefer to develop the noncommutative structure
by spelling out the choices for the commutation relations.
\medskip\par
We will now show that the full algebra of $x,\xi,p,\pi$
includes the generators of
translations in position and momentum space:
\BE
\BA{rclrcl}
T &=&-\i \ (\xi\!\cdot\! p)\ \qquad  & Tx &=&xT+\xi \\
S &=&\ \i \xp\   &  Sp&=& pS+\pi
\EA
\EE
In the following we will use $\partial =-\i p$ to avoid imaginary
units.\medskip\par

We have to compute the commutation relations of
the generators and other scalar
combinations with the quantum vectors. This can
be done most easily using the relation (6), e.g.
\BE
\renewcommand{\arraystretch}{2}
\BA{rcl}
\partial^i \xp &=& g_{jk}\ \partial^i x^j\pi^k  \\
 &=& g_{jk}g^{ij}\pi^k +
\underline{ g_{jk}q^2{\R^{-1}}^{ij}_{\ \ j\pr i\pr} }\
x^{j\pr}\partial^{i\pr}\pi^k  \\
 &=&  \pi^i + \underline{ g_{j\pr k\pr }\R^{k\pr i}_{\ \ i\pr k} }\
x^{j\pr}\partial^{i\pr}\pi^k\\
 &=&  \pi^i + g_{j\pr k\pr }\ x^{j\pr}\pi^{k\pr} \partial^i \\
 &=&  \pi^i+ \xp \partial^i
\EA
\EE
where the underlined expressions are equal due this relation.
\medskip\par
In order to iterate the relation (20) two further commutation
relations are of importance:
\BE
\renewcommand{\arraystretch}{1.8}
\BA{rcl}
x^i \xp &=& g_{jk}x^i x^j \pi^k   \\
 &=&   \underline{ g_{jk}
\left( q^2{\R^{-1}}^{ij}_{\ \ j\pr i\pr} \right } \left
+ (1-q^N){P_T}^{-1}}^{ij}_{\ \ j\pr i\pr} \right)
x^{j\pr}x^{i\pr}\pi^k \\
&=& \ds{\underline{ g_{j\pr k\pr} \R^{k\pr i}_{\ \ i\pr k}}
x^{j\pr}x^{i\pr}\pi^k +\frac{1-q^N}{\N}g_{jk}g^{ij}\xx \pi^k } \\
&=& q^2 \xp x^i +  \mu\ \xx\pi^i
\EA
\EE
and
\BE\renewcommand{\arraystretch}{1.8}
\BA{rcl}
\pi^i\xp &=&  g_{jk}\ \pi^ix^j\pi^k} \\
&=& \underline{ q^{-2} g_{jk} \R^{ij}_{\ \ j\pr i\pr} }\
x^{j\pr}\pi^{i\pr}\pi^k\\
&=& \underline{ g_{j\pr k\pr} {\R^{-1}}^{k\pr j}_{\ \ i\pr k} }\
x^{i\pr}\pi^{j\pr}\pi^k\\
&=& \ds{ g_{j\pr k\pr}
\left( \frac{1}{q^2}\1^{k\pr i}_{\ \ i\pr k}
-\frac{1-q^N}{q^2}\,{P_T}^{k\pr i}_{\ \ i\pr k}\right) \
x^{j\pr}\pi^{i\pr}\pi^{k}  } \\
&=&  q^{-2}\xp \pi^i- q^{-2}\mu\ x^i\pp
\EA
\EE
where we have used the relations (13) on symmetric coordinates.
Note that this result
depends on the right choice of $\R^{-1}$ in (16).
\medskip\par

With the above
technique we get the following relations between the quantum
vectors and the scalar combinations (the vector indices are
omitted for simplicity)
\[
\BA{rcl}
\partial  \xp &=& \xp \partial +\pi \\
\pi \xp &=& q^{-2}\xp \pi -q^{-2}\mu x\pp \\
x \xp &=& q^2\xp x +\mu\xx\pi \\
\xi \xp &=& \xp \xi \medskip\\
x\ad &=& \ad x -\xi \\
\xi\ad &=& q^2\ad\xi  +q^2\mu\ \partial\XX \\
\partial \ad &=& q^{-2}\ad\partial -\mu\dd\xi  \\
\pi\ad &=& \ad\pi
\EA
\]

\BE
\BA{rcl}
x\ap &=& q^2 \ap x -(q^2-1)\,(\xi\!\cdot\! x)\ \pi + (q^2-1)\
\xi\xp \\
\xi\ap &=& q^2\ap\xi  +\mu\XX\pi \\
\partial \ap &=& q^{-2}\ap\partial -(1-q^{-2})\ad\pi +
(1-q^{-2})\  \xi\ (\partial \!\cdot\! \pi)  \\
\pi \ap &=& q^{-2}\ap\pi -q^{-2}\mu\ \xi\pp  \medskip\\
x \xd &=& q^{-2}\xd x +\wt\mu\xx\partial -q^{-2} x \\
\xi \xd &=& \xd \xi \\
\partial  \xd &=& q^2 \xd \partial  -q^2\wt\mu\ x\dd +\partial\\
\pi \xd &=& \xd \pi
\EA
\EE

with $\mu$ and $\wt\mu$ as above.
\medskip\par

We will also need the commutation relations with the additional
scalar combinations
appearing in (24), that can be calculated accordingly.

\BE
\BA{rclrcl}
\partial \xx &=& q^2\xx \partial + (1+q^{-(N-2)})x  &  && \cr
x\dd &=& q^{-2} \dd x  -q^{-2}(1+q^{-(N-2)})\ \partial &  && \cr
x\XX &=&q^2\XX x  &
\partial \XX &=&q^{-2} \XX\partial   \cr
x\pp  &=&q^2 \pp x &
\partial \pp &=&q^{-2}\pp\partial \cr
\pi \xx &=&q^{-2} \xx\pi   &
\xi\xx &=&q^{-2}\xx\xi \cr
\pi\dd &=&q^2 \dd\pi  &
\xi\dd &=&q^2 \dd \xi \cr
\pi \XX &=&q^{-2} \XX\pi   &
\xi\pp  &=&q^2 \pp \xi
\EA
\EE
For later convenience we list also some relations among the scalars
that follow directly
from (24) and (25).

\[
\BA{rcl}
    \pp \xp  &=& q^{-2} \xp \pp \cr
    \xx \xp  &=& q^2   \xp \xx \cr
    \XX \xp  &=&        \xp \XX \medskip\cr
    \ap \xp  &=& q^{-2} \xp \ap  -q^{-2}\mu  \ax \pp \cr
    \ax \xp  &=& q^2   \xp \ax  +q^{-2}\mu  \xx \ap
\EA
\]
\BE
\BA{rcl}
    \ax \ap  &=&   q^2 \ap \ax  -q^{-2}\mu  \XX \xp \cr
    \xx \ap  &=& q^{4} \ap \xx \cr
    \pp \ap  &=& q^{-2} \ap \pp \medskip\cr
    \pp \xx  &=& q^{-4} \xx \pp \cr
    \pp \XX  &=& q^{-4} \XX \pp \cr
    \pp \ax  &=& q^{-4} \ax \pp \cr
\EA
\EE

Corresponding relations for $T=\xi\!\cdot\!\partial $ follow from
a substitution
symmetry described below (46).

\section{A construction for generalized $q$-exponentials}

In the last section we have identified the generators $T$
and $S$ for translations in
position and momentum space.
We now demonstrate how these relations can be integrated
to yield the finite displacement operators. In the classical
case and also for the quantum
plane these operators can be
given in terms of a power series in the generator, i.e. the
exponential map and Jackson's $q$-exponential, respectively.
As mentioned above in the
physically important cases
of Minkowski and Euclidean spaces the structure of the quantum
deformation is more intricate.
\medskip\par
Let us analyse the first terms of the exponential series for the
operator that shifts
$\partial $ to $\partial +\pi$ in momentum space
\BE
\exp(x,\pi)=1+\xp+\dots
\EE

Differentiating this series by $\partial^i$ should yield
the exponential multiplied with
$\pi^i$. This is clearly the case for the first term due to (15).
How has the second term to
look like? Since
\BE
\partial^i \xpn(2)=  \xpn(2)\partial^i+
(1+q^{-2})\xp \pi^i -q^{-2}\mu\,  x^i\pp
\EE
we need also terms involving $\xxpp$ with
\BE
\partial^i \xxpp = \xxpp\partial^i +
(1+q^{-(N-2)})x^i\pp = -\frac{q^2-1}{\mu}\ x^i\pp
\EE
to compensate the unwanted piece in (28).
Combining the results we find
\BE
\partial^i \left[\xpn(2)-\frac{\mu^2}{q^2(q^2-1)}\xxpp\right]
= (1+q^{-2})\xp \pi^i\ .
\EE
The differentiation $\partial $ reduces the degree in $x$
of any scalar monomial by one.
If the exponential series is a sum of such terms ordered
by their degree in $x$, then the
differential reduces each term to the preceding one, except
1 which is annihilated, multiplied
by $\pi$. The coefficients have to be determined such that
term containing $x\pp$ as in (28)
are cancelled.
{}From this result
it is already clear that --- unless $\pi\cdot\pi$
vanishes --- the Jackson exponential cannot be used
to construct the operator of finite
translations.
Iterating the procedure we can integrate the exponential series
using $[n]_{q^{-2}}=1+q^{-2}+\dots+q^{-2(n-1)}$
\BE\renewcommand{\arraystretch}{2.5}
\BA{rcl}
\exp_{q^{-2}}(x;\pi)=1&&+\xp
+\ds{\frac{1}{[2]_{q^{-2}}}\left[\xpn(2)
   -\mu^2\frac{1}{q^2(q^2-1)}\xxpp \right]  }  \\
&&+\ds{\frac{1}{[3]_{q^{-2}}}\left[\xpn(3)
      -\mu^2\,\frac{2q^2+1}{q^4(q^2-1)}\xp\xxpp \right] } \\
&&+\ds{\frac{1}{[4]_{q^{-2}}}\left[\xpn(4)
      -\mu^2\,\frac{3q^4+2q^2+1}{q^6(q^2-1)}\xpn(2)\xxpp \right } \\
&&\qquad\qquad\qquad\qquad\qquad   +\ds{\left\mu^4\,
\frac{q^4+q^2+1}{q^8(q^2-1)^2}[\xxpp]^2 \right]} \\
&&+\dots
\EA
\EE

that is defined by the property
\BE
\partial\ \ \exp(x;\pi)
 = \exp(x;\pi)
\ \ (\partial +\pi)  \ .
\EE
In order to write a closed form for the coefficients of the
additional terms
$\xxpp$ that appear
in this generalized exponential we have to consider the
following three relations coming from the basic equations
(21), (22), and (23):
\BE\renewcommand{\arraystretch}{2}
\BA{rcl}
\partial \xpn(n) &=& \ds{ \xpn(n)\partial
+ \sum^{n-1}_{k=0}\xpn(n-1-k)
\left\{ \pi\xpn(k) \right\}  } \smallskip\cr
\pi \xpn(k)&=&
\ds{ q^{-2k}\left( \xpn(k)\pi
-\mu\sum^{k-1}_{l=0} \xpn(k-1-l)\left[ x\xpn(l) \right] \pp
\right) }   \smallskip\cr
x \xpn(l) &=&  \ds{ q^{2l}\left( \xpn(l) x
+\mu\sum^{l-1}_{m=0} \xpn(l-1-m)\left\{ \pi \xpn(m) \right\}
\xx \right) }
\EA
\EE

We can observe that the relations iterate in two steps $\{...\}\to
[...]\to \{...\}$ etc.
reducing the power in $\xp$ at least by one at each step.\medskip\par

We introduce $q$-combinatorial coefficients
\BE
\BA{rcl}
C^{(n)}_f &=& \ds{
\sum_{k_1=0}^{n-1}q^{-2k_1}
\sum_{k_2=0}^{k_1-1}q^{2k_2}
\sum_{k_3=0}^{k_2-1}q^{-2k_3}  \dots
\sum_{k_f=0}^{k_{f-1}-1}q^{\pm2k_f}  }
\cr
&&\cr
&=& \ds{
\sum_{n> k_1> k_2...> k_f\geq 0}
q^{-2(k_1-k_2+-\dots \pm k_f)}  }
\EA
\EE
that are $q$-deformed binomial coefficients,
$C^{(n)}_f={n\choose f}$ for $q=1$.
Hence there exists also the following recursive definition
\BE\renewcommand{\arraystretch}{1.2}
C^{(n)}_f=\left\{
\BA{ccccl}
q^{-2}\,C^{(n-1)}_{f-1}  &+& C^{(n-1)}_f && \hbox{for $f$ even} \\
C^{(n-1)}_{f-1} &+&  q^{-2}\,C^{(n-1)}_{f} && \hbox{for $f$ odd}
\EA
\right
\EE
which can be proved immediately by splitting the sum in (34)
\BE
\sum_{n> k_1> k_2...> k_f\geq 0} =
\sum_{n> k_1> k_2...> k_f= 0} +
\sum_{n> k_1> k_2...> k_f> 0}
\EE
and shifting the variables by one $k_i\to k_i-1$ changes the argument
of the sums by a
factor $q^{-2}$ if its number is odd. Using these coefficients we find
\BE
\BA{rcl}
\partial \xpn(n) = \xpn(n)  \partial
&+& {\ds
\sum_{f=1,3...\leq n} (-\mu^2)^{\frac{f-1}{2}}
C^{(n)}_f
\xpn(n-f)\pi\ [\xxpp]^{\frac{f-1}{2}}
} \cr
&-& {\ds
\sum_{f=2,4...\leq n} \mu(-\mu^2)^{\frac{f-2}{2}}
C^{(n)}_f
\xpn(n-f)x\pp\ [\xxpp]^{\frac{f-2}{2}}
}
\EA
\EE
The coefficients $C^{(n)}_{f}$ can be written in terms of
$q$-numbers, e.g. $C^{(n)}_{0}=1$, $C^{(n)}_{1}=[n]_{q^{-2}}$,
$C^{(n)}_{2}=([n]_{q^{-2}}-n)/(1-q^2)$ etc.
For later use define also $\wt C^{(n)}_f$ as
$C^{(n)}_f$ with $q^{-1}$ replacing $q$.
In order to account for the additional terms
due to the trace projector in (13) we
have to choose the $n$th term in the expansion
of the $q$-exponential function to be
 of the form
\BE
(x;\pi)^{(n)} = \xpn(n)+\sum_{a=1,2...\leq n/2}\nu^{(n)}_a
\xpn(n-2a)[\xxpp]^{a}
\EE
One can now consider two cases in which the derivative reduces
this term in the
expansion to the preceding one:
\BE\BA{lrcl}
(i) \quad & \partial (x;\pi)^{(n)} &=& (x;\pi)^{(n)}\partial
+[n]_{q^{-2}} (x;\pi)^{(n-1)}\,\pi \cr
(ii) & \partial \wt{(x;\pi)}^{(n)} &=& \wt{(x;\pi)}^{(n)}\partial
+[n]_{q^{2}}\ \pi\ \wt{(x;\pi)}^{(n-1)}
\EA
\EE
where $\wt{(x;\pi)}$ is defined by (38) with $\wt\nu$ instead of $\nu$.
The coefficients $\nu^{(n)}_s$ are determined uniquely by each of these
requirements.
For the case ($i$) we derive in the appendix
\BE
\nu^{(n)}_a =
(-\mu^2)^a
\sum_{l=0}^{a-1}
\left(\frac{1}{q^2-1}\right)^{l+1}
\sum_{a> k_1>...k_l>0}
\frac{C_{2(a-k_1)}^{(n-2k_1)}}{[a]_{q^2}}
\frac{C_{2(k_1-k_2)}^{(n-2k_2)}}{[k_1]_{q^2}}
\dots
\frac{C_{2(k_{l-1}-k_l)}^{(n-2k_l)}}{[k_{l-1}]_{q^2}}
\frac{C_{2k_{l}}^{(n)}}{[k_l]_{q^2}}
\EE
as for the case $(ii)$
\BE
\wt\nu^{(n)}_a =
(-\mu^2)^a
\sum_{l=0}^{a-1}
\left(\frac{q^{-4}}{q^{-2}-1}\right)^{l+1}
\sum_{a> k_1>...k_l>0}
\frac{\wt C_{2(a-k_1)}^{(n-2k_1)}}{[a]_{q^{-2}}}
\frac{\wt C_{2(k_1-k_2)}^{(n-2k_2)}}{[k_1]_{q^{-2}}}
\dots
\frac{\wt C_{2(k_{l-1}-k_l)}^{(n-2k_l)}}{[k_{l-1}]_{q^{-2}}}
\frac{\wt C_{2k_{l}}^{(n)}}{[k_l]_{q^{-2}}}
\EE
Note for $l=0$ there is no summation but still a term proportional
to $C^{(n)}_{2a}$.
\medskip\par

We can now define generalized exponentials for both cases
\BE\BA{rcl}
\exp_{q^{-2}}(x;\pi) &=& \ds{
\sum_{n=0}^{\infty}\frac{(x;\pi)^{(n)}}{[n]_{q^{-2}}!} } \medskip\cr
\wt{\exp}_{q^{2}}(x;\pi) &=& \ds{ \sum_{n=0}^{\infty}
\frac{\wt{(x;\pi)}^{(n)}}{[n]_{q^{2}}!} } \EA
\EE

Due to their construction the generalized $q$-exponentials have the
following
characteristic properties
\BE\BA{rcl}
\partial\ \ \exp_{q^{-2}}(x;\pi)
 &=& \exp_{q^{-2}}(x;\pi)
\ \ (\partial +\pi)  \cr
\wt{\exp}_{q^{2}}(x;\pi)\ \ \partial
 &=& (\partial -\pi)\ \ \wt{\exp}_{q^{2}}(x;\pi)
\EA
\EE
More general for $p$ we can write
\BE\BA{rcl}
f(p,\xi)\ \exp_{q^{-2}}(x;\i\pi)
&=& \exp_{q^{-2}}(x;\i\pi)\ f(p+\pi,\xi )  \cr
\wt{\exp}_{q^{2}}(x;\i\pi)\ f(p,\xi)
&=& f(p-\pi,\xi )\ \wt{\exp}_{q^{2}}(x;\i\pi)
\EA
\EE
The function $f$ can depend arbitrarily of
$\xi$ also due to the fact that $\xi$
commutes with both $(x\!\cdot\!\pi)$ and
$(x\!\cdot\!x)(\pi\!\cdot\!\pi)$ and hence
with the whole exponential that depends
only on the scalar combinations $\xp$,
$\pp$ and $(x\!\cdot\!x$), i.e. the
structure of the exponential function in this
$q$-deformed generalization depends on the length
of the translation. Clearly, the
exponentials built from scalar
combinations are invariant under quantum orthogonal
or Lorentz transformations. \par
It is a simple consequence of
relations (43) that the two exponentials are the inverses
of one another for negative argument:
\BE
\exp_{q^{-2}}(x;\pi)\ \wt{\exp}_{q^{2}}(-x;\pi) =1
\EE
{\it Proof:} From (43) it is clear that this combination
commutes with the derivative and
hence is independent of $x$; since $x$ and $\pi$ appear
in the definition of the exponentials
always with the same power it is also independent of
$\pi$ and consequently it must be the
number appearing in the product which is 1. Due to
the fact that
$x\!\cdot\!\pi$ and $(x\!\cdot\!x)(\pi\!\cdot\!\pi)$
commute and thus all $(x;\pi)^{(n)}$,
the exponentials are true inverses.
Note that the statement remains true for replacing
$x$ with
$\xi$ or $x+\xi$ since the $x$-$\pi$ commutation
relations coincide with those of $\xi$-$\pi$.
\medskip\par

In order to obtain the translation generator for the momentum
space one can make use of a symmetry of the
commutation relations (24). The  relations used
in momentum space are mapped to those of the
position space by following substitution:
\BE
\left\{
\BA{rcl}
i)&&\hbox{reverse order}\cr
ii)&&\partial  \to x \ ,\quad\pi \to \xi \cr
   &&\hbox{i.e.\ }\xp \to \ad       \cr
   &&\hbox{\phantom{i.e.\ }}\pp\to \XX  \ ,\quad  \xx\to \dd    \cr
\ \ iii)&& q\to q  \ ,\quad \hbox{i.e.}\quad   \mu \to \mu
\EA
\right
\EE
Thus from (43) we get immediately
\BE\BA{rcl}
x\ \ \wt{\exp}_{q^{2}}(\xi;\partial)
 &=& \wt{\exp}_{q^{2}}(\xi;\partial)
\ (x -\xi)  \cr
\exp_{q^{-2}}(\xi;\partial)\ \ x
 &=& (x +\xi)\ \exp_{q^{-2}}(\xi;\partial)
\EA
\EE
or for polynomials in $x$ and $\pi$
\BE\BA{rcl}
f(x,\pi)\ \wt{\exp}_{q^{2}}(\xi;-\i p)
&=& \wt{\exp}_{q^{2}}(\xi;-\i p)\ f(x-\xi,\pi )  \cr
\exp_{q^{-2}}(\xi;-\i p)\ f(x,\pi)
&=& f(x+\xi,\pi )\ \exp_{q^{-2}}(\xi;-\i p)
\EA
\EE
This exponential depends only on the scalar combinations $\ap$,
$\XX$ and $\dd$.
The inverse properties are accordingly.
\medskip\par

Having defined the generalized exponentials by their
differentiation properties
it is an interesting feature that they satisfies also
an addition theorem, e.g.\ for
position variables
\BE
\exp_{q^{-2}}(x+\xi;\pi)=
\exp_{q^{-2}}(x;\pi)\ \exp_{q^{-2}}(\xi;\pi)\ .
\EE
{\it Proof:} Considering the product $\exp(\xi;\partial )\
\exp(x;\pi)$ there are two
 ways to commute the exponentials using (44) or (48). Either
the first one is used as
translation of $x$ or the second one as translation in
$\partial $. As a consequence
we find
\BE
\exp_{q^{-2}}(x+\xi;\pi )\ \exp_{q^{-2}}(\xi;\partial ) =
\exp_{q^{-2}}(x;\pi)\ \exp_{q^{-2}}(\xi;\partial +\pi)
\EE
Next, with (38) we
observe\footnote{This observation is due to Chong-Sun Chu.} that
\BE
\exp_{q^{-2}}(\xi;\partial +\pi)=
\exp_{q^{-2}}(\xi;\pi) +
(\hbox{terms containing $\partial,\ \xi $ and partly $\pi$})
\EE
and due to the fact that all commutation relations among
$\partial ,\ \xi$ and $\pi$
are homogeneous, no additional term independent of
$\partial$ can arise besides
$\exp_{q^{-2}}(\xi;\pi)$ by using these relations
to order the expressions. For
$\exp_{q^{-2}}(\xi;\partial )$ the only term
independent of $\partial $ is 1.
In (52) $x$ is already ordered left of
$\partial $. To be an identity the terms
independent of $\partial $ in the relation
(52) must coincide on both sides yielding
the desired result.

Since (49) is an algebraic statement for $x,\ \xi$ and $\pi$,
it holds for all
triples of quantum vectors that have the same commutation relations.

\section{Eigenstates and Fourier transformations}

As introduced in \cite{CZ} for the case of the quantum plane,
we can now use position
and momentum vacua to construct eigenstates of the momentum
operator $p$ and the
position operator $x$ with (noncommutative) eigenvalues $\pi$
and $\xi$. We use
again $\partial =-\i p$ to avoid imaginary units.\medskip\par
Let us define zero-momentum or `vacuum' states:
\BE\BA{rcl}
|0\rangle &=& |\pi=0\rangle\qquad\hbox{with}\qquad \partial
|0\rangle =0\cr
\langle 0| &=& \langle \pi=0|\qquad\hbox{\phantom{with}}\qquad
\langle 0|\partial  =0
\EA
\EE
The eigenvalues $\pi$ `commute' with  the vacuum $\pi|0\rangle
=|0\rangle\pi$ since $\partial \pi|0\rangle=\R^{-1}\pi
\partial   |0\rangle=0$,
i.e. the $\pi$ in front of the vacuum does not spoil its
property (54) and hence
one can write it on the r.h.s. of the ket.\par

Due to relation (43) we can define arbitrary momentum
eigenstates using the
vacuum and the exponential:
\BE\BA{rclcrcl}
|\pi\rangle &=& {\exp}(x;\pi)|0\rangle
&\hbox{\quad with \quad}&
\partial\ |\pi\rangle &=& |\pi\rangle\ \pi \cr
\langle \pi| &=&\langle 0|{\wt\exp}(-x;\pi)
&\hbox{\quad with \quad}&
\langle \pi|\partial &=&\pi\,\langle \pi|
\EA\EE
This definition of the states has the favourable property that
\BE\BA{rcl}
f(\partial)\ |\pi\rangle &=& |\pi\rangle\ f(\pi) \cr
\langle \pi|\ f(\partial) &=&f(\pi)\ \langle \pi|
\EA
\EE
Had we chosen a different definition of the states such that
the eigenvalues appeared
on the same side of the
state as the operator, the evaluation of a function on such a
state would have changed
the function itself. Assuming that the vacuum state is normalizable,
and hence its normalization
should be unity, we can write in particular
\BE
\langle \pi|\,f(\partial)\,|\pi\rangle &=&  f(\pi) \ .
\EE
This shows that analog to the undeformed case the
operators $\partial $ and $x$ are
to be applied inside of brackets of states yielding
results in term of their eigenvalues.
\medskip\par
Applying (49) or (52) on the momentum vacuum we get
\BE
\exp_{q^{-2}}(x+\xi;\pi)|0\rangle =
\exp_{q^{-2}}(x;\pi)|0\rangle\ \exp_{q^{-2}}(\xi;\pi)
\EE
and
\BE
\langle 0|\wt{\exp}_{q^{2}}(x+\xi;\pi) =
\wt{\exp}_{q^{2}}(\xi;\pi)\ \langle 0|\wt{\exp}_{q^{2}}(x;\pi)
\EE
using the inverse relation (45).
{}From this it is clear how momentum eigenstates transform under
a finite translation,
they acquire a
noncommutative phase:
\BE\BA{rcl}
x\to x+\xi\ :\qquad &&
|\pi\rangle\to |\pi\rangle \ {\exp}_{q^{-2}}(\xi;\pi)\cr
&& \langle \pi|\to \wt{\exp}_{q^{2}}(-\xi;\pi)\ \langle \pi|
\EA\EE
Introducing $\xi\pr$ such that $(\xi\pr,\xi,\pi)$ has the same
noncommutative structure as
$(x,\xi,\pi)$, we find, that the transformations form a kind of
noncommutative
translation group:
\BE\BA{rcl}
x\to x+\xi+\xi\pr\ :\qquad && |\pi\rangle\to |\pi\rangle \
{\exp}_{q^{-2}}(\xi\pr;\pi){\exp}_{q^{-2}}(\xi;\pi)\cr
&=& |\pi\rangle\to |\pi\rangle \ {\exp}_{q^{-2}}(\xi\pr+\xi;\pi)
\EA\EE
\medskip\par

One can easily see that the states introduced are of equal norm
\BE
\langle \pi|\pi\rangle=
\langle 0|\wt{\exp}_{q^{2}}(-x;\pi)\exp_{q^{-2}}(x;\pi)|0\rangle
=\langle 0|0\rangle
\EE
and this norm is also preserved under translations since the
noncommutative
phases cancel.\medskip\par

Orthogonality relations of states are in the noncommutative
setting more complicated.
It is easy to recognize that
the vacuum state is still orthogonal to any other
momentum state
\BE
\langle 0|\pi\rangle\pi=\langle 0|\partial |\pi\rangle=0\quad
\hbox{i.e.}\quad \langle 0|\pi\rangle=0\ .
\EE

To analyse the situation for two different momentum states we
have to introduce another
quantum plane $\pi\pr$
\BE
\pi\pi\pr=\R \pi\pr\pi\qquad \pi\pr x=q^{-2}\R x\pi\pr
\EE
i.e.\ the $\pi$-$\pi\pr$ relations are the same as the
$\pi$-$\partial $ relations
and  otherwise $\pi\pr$ commutes like $\pi$.
In particular we find
\BE
[\pi\pr,\exp_{q^{-2}}(x;\pi)]=0\qquad\hbox{or}\qquad
\pi\pr|\pi\rangle =
|\pi\rangle \pi\pr\ .
\EE
However, for those kinds of states defined above it can only be
inferred that
\BE
\pi\pr\langle \pi\pr|\pi\rangle =
\langle \pi\pr|\partial |\pi\rangle=
\langle \pi\pr|\pi\rangle\pi\ .
\EE
For the case that $\pi$
or $\pi\pr$ commuted with the matrix element one could
conclude that
\BE
\langle \pi\pr|\pi\rangle=0\qquad
\hbox{iff}\qquad\pi\pr\neq\pi\ .
\EE
This, however, may not be the case
in general.\footnote{In such a case we could introduce
a different set of bra states $ ^\#\langle \pi| =\langle
0|{\exp}(-x;\pi)$
with $ ^\#\langle \pi|\partial
=\langle \pi|\,\pi $ and would then find in general
$ ^\#\langle \pi\pr|\pi\rangle=0$ iff
$\pi\pr\neq\pi $ due to (63).}

\bigskip
The basic idea to formulate Fourier
transformations of noncommutative spaces is
that there exists an analog of integration
as a linear map from the functions in
the quantum variables to the complex numbers
\cite{WZ,CZ} or to a noncommutative
space again \cite{KM}. The integral is related
to the states we have introduced
since it has to be a linear functional that vanishes
for a derivative
\BE
\langle \partial ...\rangle =0
\EE
which is reminiscent of
$\langle 0|\partial =0$ and expresses the translational
invariance of the integral.\par

Let us complement the momentum states $|\pi\rangle_p$ with
position states
$|\xi\rangle_x$
with the obvious analog relations.
It is a natural assumption that a general state
$|\hbox{{\bf f}}\,\rangle$ can either be written as a function in
$x$ on the momentum vacuum or as a function in $p$ on the
position vacuum.
\BE
|\hbox{{\bf f}}\,\rangle
= f(x)\,|0 \rangle_p
= \wt{f}(p)\,|0 \rangle_x
\EE
This gives on the one hand the relations
\BE
\BA{rcl}
_p\langle \pi | \hbox{{\bf f}}\, \rangle
&=&
_p\langle 0 | \wt{\exp}_{q^{2}}(x;-\i \pi)\wt{f}(p) | 0\rangle_x   \\
&=&
_p\langle 0 | \wt{f}(p+\pi) \wt{\exp}_{q^{2}}(x;-\i \pi)| 0\rangle_x
\\ &=&
\wt{f}(\pi)\ _p\langle 0 | 0\rangle_x \medskip\\
_x\langle \xi | \hbox{{\bf f}}\, \rangle
&=&
_x\langle 0 | {\exp}(\xi;\i  p){f}(x) | 0\rangle_p   \\
&=&
_x\langle 0 | {f}(x+\xi) {\exp}(\xi;\i  p)| 0\rangle_p  \\
&=&
{f}(\xi)\ _x\langle 0 | 0\rangle_p
\EA
\EE
where we assume that the vacua are normalized to
unity as above. With this relation
we have found the analog to the notion of position
and momentum representation of an
abstract state vector in quantum mechanics. On the
other hand the Fourier
transformations can now be stated as
\BE
\BA{rcccccccl}
\wt{f}(\pi)&=&({\cal F}f)(\pi)
&=&
_p\langle \pi | \hbox{{\bf f}}\, \rangle
&=&
_p\langle 0 | \wt{\exp}_{q^{2}}(x;-\i  \pi){f}(x) | 0\rangle_p
&=&
\langle \wt{\exp}_{q^{2}}(x;-\i  \pi){f}(x) \rangle
\\
f(\xi)&=&({\cal F}^{-1}\wt{f})(\xi)
&=&
_x\langle \xi | \hbox{{\bf f}}\, \rangle
&=&
_x\langle 0 | {\exp}(\xi;\i  p)\wt{f}(p) | 0\rangle_x
&=&
\langle {\exp}(\xi;\i  p)\wt{f}(p) \rangle
\EA
\EE
where it is understood that $x$-variables
inside momentum vacua and $p$-variables
inside position vacua are integrated over.

\newpage

\section{Conclusion and Discussion}

As we have argued in the introduction
the quantum deformation of spacetime
is some natural extension of the idea
of quantization. It is the translation
of a scheme well-known for the angular
momentum in quantum mechanics to
spacetime and momentum.
This approach develops a noncommutative
geometry for spacetime from the point
of view more closely related to quantization
than the constructions pioneered
by Connes. There one  adds to the usual
spacetime manifold a noncommutative
internal space of two points yielding a
two-sheeted `universe' that is more
geometrically motivated \cite{Connes}.
However, there is obviously a common
goal in bridging the gap between geometrical
and quantum concepts that already
meet at the purely algebraic level.

The main application of the results obtained
in this article are relativistic
particle states in quantized spacetime.
As the Dirac equation can be easily
formulated in this setting \cite{Cam}
(for higher spin see \cite{Pillin}) and
solved in momentum space \cite{Diss} the
states and Fourier transformations
constructed in section 4 allow to give
the free solutions in position space.
Their transformation properties under
translations have been determined and
turn out to be noncommutative phases.
These results may be seen as an alternative
approach to the problem of finding an
appropriate quantum Poincar\'e symmetry. In
a next step one can now introduce an analog
of electromagnetic interactions in the
$q$-Dirac theory which is most easily done
in the position space representation
\cite{prep}.\medskip\par
 One should note that, although we have treated
momentum and position space always
equally in this article, they also have different
properties. First, it is the momentum
generators that give the quantum symmetry of
translational invariance and second it
is not possible to have linear conjugation
structures on both momentum and position
space in the orthogonal case \cite{Ogi}.
{}From the point of view of particle states
it seems most natural to adopt a linear
conjugation structure for the momentum
operators \cite{Cam}.
In this case we find two sets of position
operators $x$ and $\ol{x}$ (which
is a well-known property of discrete
lattice structures) which have to be
combined in a nontrivial way for the
identification of a hermitean position
observable. For the one-dimensional
$q$-deformed Heisenberg algebra cf. \cite{Schwenk}.
\medskip\par

The essential feature in the quantum
group approach is that the replacing operator
algebra is a {\it deformation} of the commutative
case. This means that if one starts
with all products of the variables involved (free
algebra) the number of commutation
relations identifying products of different ordering
stays the same, i.e. the Poincar\'e
series does not change under the deformation.
In \cite{Maj} it was attempted to give exponential
series for any $\R$-matrix. The
(braided) construction involved inverting $\R$-matrix
combinations as $\1+\R$ coming
from computing $\partial xx...$ using the $\R$-matrix
relation in general.
 These combinations are for all cases where a consistent
linear differential
calculus is assumed necessarily projectors and hence not
invertible.\footnote{There is a bit of confusion about
this in \cite{KM}
as the Hecke case
is cited as an
`example'. $\1+\R$ in the correct normalization
required by (9) is clearly not invertible in this
case and it is mysterious how
one can ``know without any computation'' that this
reduces to the $q$-numbers.}
In order not to change the construction the $q$-deformed
Minkowski space was
sacrificed for a free algebra that contains 16 instead of
10 linearly independent
quadratic elements. In addition it becomes unclear why one
should use $\R$ matrices
for the commutation between different quantum spaces, since
no Yang-Baxter-Relation
has anymore to hold for one matrix, less restrictive relations
apply in this case.
As a consequence no Fourier transformations for {\it q-deformed }
Minkowski or
 Euclidean spaces can be defined by this construction \cite{KM}.
As pointed out in section 3 for vanishing momentum square $\pp$
the additional
pieces in the exponential series disappear and thus it reduces
to the Jackson
exponential. There is no need to introduce new exponentials as
suggested in \cite{Meyer}.
\medskip\par

Let us conclude with a comment
on a confusion of tongues. What is meant by the terms
quantum deformation of Lorentz
group and Minkowski space? In the spirit of the
inventors of the theory we can
find deformations {\it in a strong sense}: The
quantum Minkowski space is a
deformation (of the same Poincar\'e series) with
a central quadratic element,
its `length'. The quantum Lorentz group transforms
this space `covariantly' which
means in particular preserving its noncommutative
structure and leaving the length
invariant and thus being a {\it quantum symmetry}.
As we have seen particle states of
arbitrary mass can be constructed in this framework.
We also noticed that they have a well-defined
behaviour under translations even when
the question of a quantum deformations of the
Poincar\'e group in the strong sense has
not yet been answered fully. (But cf. \cite{WP}.)\par
One the other hand there are quantum deformations
of Lorentz and Poincar\'e
groups {\it in a weak sense} that only require that
there exists a limit in
which the usual structure is recovered, but there need
not be analogs of invariants
and quantum symmetries in the deformed case. If the
Minkowski length is not invariant
massive particle states cannot be
constructed. If one restricts oneself to the massless
(high-energy limit) case, however,
one might then rather consider quantum deformations
of the conformal
group than Lorentz and Poincar\'e groups, that in most cases
are not quantum subgroups of the latter.

\subsection*{Acknowledgements}

The author would like to thank Bruno Zumino
for drawing attention to the topic
of this article and for fruitful exchange and
Chong-Sun Chu for discussions
and reading of the manuscript. \par
{\small
This work was supported
by the Fritz Thyssen Foundation and
by part by the Director,
Office of Energy Research, Office of High Energy and
Nuclear Physics, Division of High Energy Physics of the
U.\ S.\ Department of Energy under Contract
DE-AC03-76SF00098 and in part by the National Science
Foundation under grant PHY90-21139.}

\newpage

\section*{Appendix}

To determine the coefficients $\nu$ and $\wt\nu$
consider the relations (39). The appearing $q$-numbers are
already determined by $C^{(n)}_1$ and $\wt
C^{(n)}_1$. Let us work out the case $(i)$:
The derivative on $\xpn(n)$ has been calculated in (37).
In addition we need
\BE
\partial [\xxpp]^{a} = [\xxpp]^{a}\partial +[a]_{q^{2}}
(1+q^{-(N-2)})\ x\pp
[\xxpp]^{a-1}
\EE
and combining the results we find
\BE\renewcommand{\arraystretch}{1.7}
\BA{rl}
\partial \xpn(n-2a)&[\xxpp]^{a}
= \xpn(n-2a)[\xxpp]^{a} \partial
\cr
+&
{\ds
\xpn(n-2a)[a]_{q^{2}} (1+q^{-(N-2)})\ x\pp [\xxpp]^{a-1}
} \cr
+&
{\ds
\sum_{g=1,3...\leq n-2a} (-\mu^2)^{\frac{g-1}{2}}
C^{(n-2a)}_g
\xpn(n-2a-g)\pi\ [\xxpp]^{a+\frac{g-1}{2}}
} \cr
-& {\ds
\sum_{g=2,4...\leq n-2a} \mu(-\mu^2)^{\frac{g-2}{2}}
C^{(n-2a)}_g
\xpn(n-2a-g)x\pp\ [\xxpp]^{a+\frac{g-2}{2}}
}
\EA
\EE

We have to demand that the coefficient of $\xpn(n-2s)x\pp[\xxpp]^{s-1}$
that in the relation (39) case ($i$) appears after  inserting (38), and
using (37) and (70) vanishes for general $s$:
\BE
(1+q^{-(N-2)})[s]_{q^2}\nu^{(n)}_s -\mu (-\mu^2)^{s-1} C^{(n)}_{2s}
-\sum_{a=1}^{s-1}\mu(-\mu^2)^{s-1-a}
C^{(n-2a)}_{2(s-a)}\nu^{(n)}_{a}=0
\EE
i.e.
\BE
\nu^{(n)}_s= \frac{\mu(-\mu^2)^{s-1}}{(1+q^{-(N-2)})[s]_{q^2}}
\left( C^{(n)}_{2s} +\sum_{a=1}^{s-1}
(-\mu^2)^{-a}C^{(n-2a)}_{2(s-a)}\nu^{(n)}_{a}\right)
\EE
For the case $(ii)$ one can calculate accordingly determining
the coefficient
in  $[(\pi\!\cdot\!\pi)(x\!\cdot\! x)]^{s-1}\pp x\xpn(n-2s)$,
i.e. the same
combination as above in the opposite order. Note that
$(x\!\cdot\!\pi)$ and
$(x\!\cdot\!x)(\pi\!\cdot\!\pi)$
commute and hence (38) need not be changed.
The calculation is facilitated by the following observation:

In the previous calculations we ordered the $\xp$ terms
to the left hand side.
To obtain the result for the opposite order one can exploit
a formal symmetry
in the commutation relations employed: Considering $q$ and
$\mu$ as independent
(they are {\it not}!) one can observe, that the inversion
of the order of the
factors in the equations (22) and (23) and consequently
the according iterated
relations (33) gives again
correct relations if $q$ is replaced by $q^{-1}$
and $\mu$ by $-\mu$.
This gives
\BE
\wt{\nu}^{(n)}_s= \frac{-\mu(-\mu^2)^{s-1}}{(1+q^{-(N-2)})
[s]_{q^2}q^{-2s}}
\left( \wt{C}^{(n)}_{2s} +\sum_{a=1}^{s-1}
(-\mu^2)^{-a}\wt{C}^{(n-2a)}_{2(s-a)}\wt{\nu}^{(n)}_{a}\right)
\EE
where the additional factor $q^{-2s}$ appears since we do not change
$\xxpp$ in (38) and the $q$-number is not altered since it comes from
the $\partial \xp$ relation.
Iterating these relations and substituting $1+q^{-(N-2)}$ by
$-\mu/(q^2-1)$ finally yields the coefficients of the
additional terms in the expansion of
the generalized exponential as given in (42) and (43). \medskip\par
To complete the derivation we have to show in addition
that all the terms of type $\xpn(n-2s+1)\pi\,[\xxpp]^{s-1}$ cancel.
This amounts to the following consistency
condition for the case $(i)$ (for $(ii)$ accordingly)
\BE
\nu^{(n-1)}_{s-1}=\frac{(-\mu^2)^{s-1}}{q^{2(s-1)}[n]_{q^{-2}}}
\left( C^{(n)}_{2s-1}+\sum^{s-1}_{a=1}(-\mu^2)^{-a}
C^{(n-2a)}_{2(s-a)-1} \nu^{(n)}_a
\right) \ .
\EE
Without a developed $q$-combinatorics at hand for our case
the direct verification
by substitution of the above result is rather cumbersome.
So let us only give the
outline of the demonstration:
We want to prove a relation of type
\BE
\nu^{n-1}_t = \sum^t_{b=0} \be^{n,t}_b\ \nu^n_b
\EE
for all $n$ and $t$ with coefficients $\be$ that
can be read off from (75);
$\nu^n_0:=1$. We use the recursive definition (73)
i.e. with appropriate $\al$'s
\BE
\nu^{n}_s = \sum^{s-1}_{a=0} \al^{n,s}_a\ \nu^n_a
\EE
and induction over $s$. It is easy to see that
\BE
\nu^{n-1}_1 = \be^{n,1}_1 \nu^n_1 + \be^{n,1}_0
\EE
which by substituting $\nu^n_1=-\mu^2/(q^2-1)\, C^n_2$ is equivalent
to
\BE
(q^2-1)C^n_3 = q^2C^n_1C^{n-1}_2 - C^n_2C^{n-2}_1
\EE
is true using the definition of $C^n_f$. Now one has to show that
provided (76) is true for all $t<m$ it also holds for $t=m$. This
can be achieved by inserting (77) into (76) and expressing the case
$t=m$ in terms of relations for $t<m$ and interchanging the sums
appropriately.

\newpage


\end{document}